\documentclass{ws-procs975x65}

\begin{document}

\title{Multimodality of rich clusters from the SDSS DR8 
\\ within the supercluster-void network}

\author{M. EINASTO$^*$}

\address{Tartu Observatory,\\
61602 T\~oravere, Estonia\\
$^*$E-mail: maret@aai.ee\\
}

\begin{abstract}
We study the relations between the multimodality of galaxy clusters drawn from 
the SDSS DR8 and the environment where they reside. We find that multimodal 
clusters reside in higher density environment than unimodal clusters. 
We determine morphological types of superclusters and show that 
clusters in superclusters of spider morphology 
have higher probabilities to have substructure and larger peculiar velocities of 
their main galaxies than clusters in superclusters of filament morphology. Our 
study shows the importance of the role of superclusters as high density 
environment which affects  the properties of galaxy systems in them.
\end{abstract}


\bodymatter

\section{Introduction}\label{aba:sec1}
To understand the formation, evolution, and
present-day properties of the Universe we
should study the properties of structures forming the cosmic web - 
superclusters and galaxy groups and clusters in them
together. We analyse the 
properties of clusters to determine substructures in them, study the 
relations between the multimodality of rich clusters from the SDSS DR8 and the 
environment where they reside, and
compare the properties of clusters in superclusters of different morphology. 

We use data from the 8th data release of the 
Sloan Digital Sky Survey, in total 576493 galaxies from MAIN sample. Groups of galaxies are found 
with the Friends-of-Friends method, see \cite{2012A&A...540A.106T} for details. 
Data about 109 clusters with at least 50 member galaxies have been used. 
We apply the 
luminosity density field to determine superclusters of galaxies 
\cite{2012A&A...540A.106T, 2012A&A...539A..80L}, and find host superclusters
for all clusters. 

\section{Methods and results}

\begin{figure}
\begin{center}
\psfig{file=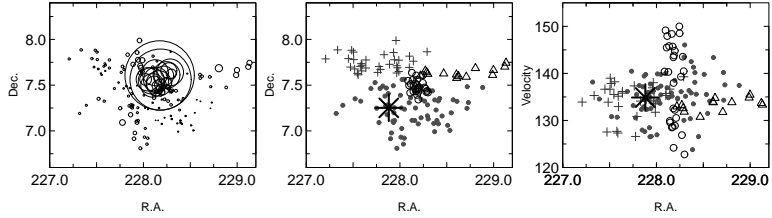, width=4in}
\caption{
The cluster 34726 (Abell cluster A2028).
From left to right: the DS test bubble plot (symbol sizes are proportional 
to the probability to have substructure), 
and R.A. vs. Dec., and R.A. vs. velocity ($10^{2}km/sec$) plots; 
the symbols show different components as found with normal mixture modelling. 
The star shows the location of the main galaxy.
}
\label{fig:grGnonG}
\end{center}
\end{figure}

We apply several 3D, 2D, and 1D tests to analyse the presence of substructure in 
clusters, the location of their main galaxies, and the velocity distribution of 
galaxies in clusters, and find that more than 80\% of clusters from our sample 
demonstrate signs of multimodality (see also \cite{2012A&A...540A.123E}). The 
main galaxies of clusters are typically located near the center of one of the 
components in the cluster, but not always in the central component. In 
Figure~\ref{fig:grGnonG} we show the results of the two 3D test: the 3D normal 
mixture modelling and the Dressler-Shectman (DS) test. 

\begin{figure}
\begin{center}
\psfig{file=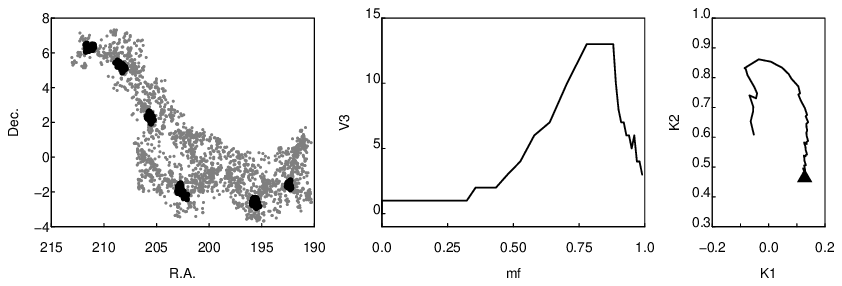, width=4in}
\psfig{file=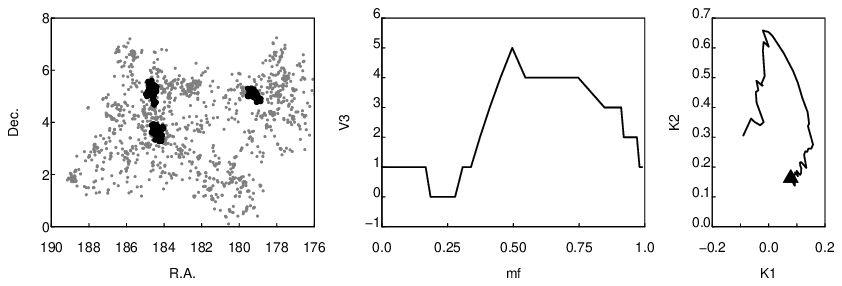, width=4in}
\caption{
Sky distribution of galaxies (left panel), the 
fourth Minkowski functional $V_3$ (middle panel) 
and the shapefinder's $K_1$-$K_2$ plane (right panel) for a
supercluster of filament morphology, SCl~027 (upper row), 
and of spider morphology, SCl~019 (lower row), two
rich superclusters in the Sloan Great Wall.
Black filled circles denote 
galaxies in clusters with at least 50 member galaxies, grey dots denote other
galaxies. Mass fraction increases
anti-clockwise along the $K_1$-$K_2$ curve. 
}
\label{fig:fil}
\end{center}
\end{figure}

We 
determined morphological types of all superclusters hosting rich clusters with 
Minkowski functionals. Using first three Minkowski 
functionals we defined shapefinders which describe the planarity (shapefinder 
$K_1$) and filamentarity (shapefinder $K_2$) of superclusters. The maximum value 
of the fourth Minkowski functional $V_3$ (the clumpiness) characterises the 
inner structure of the superclusters \cite{2007A&A...476..697E}. Superclusters 
show two main morphological types: spiders and filaments. Figure~\ref{fig:fil}
shows the fourth Minkowski functional $V_3$ vs. mass fraction 
$mf$ and the shapefinders $K_1$ and $K_2$ (morphological signature) 
for a supercluster of filament 
morphology, SCl~027, and of spider morphology, SCl~019 
\cite{2007A&A...476..697E, 2011ApJ...736...51E}. 

\begin{figure}
\begin{center}
\psfig{file=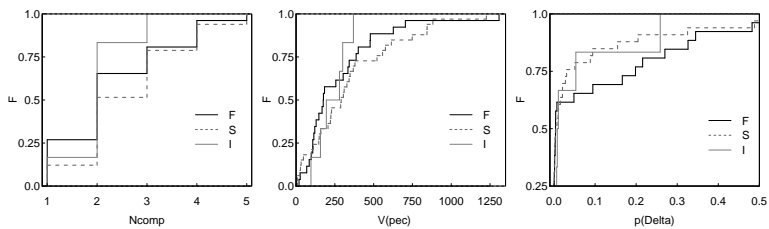, width=4in}
\caption{
Cumulative distributions of the numbers of components in clusters, $N_{\mathrm{comp}}$, 
peculiar velocities of cluster main galaxies, $V_{\mathrm{pec}}$  {\bf
(in $km~s^{-1}$)}, and 
p-value of the DS test, $p_{\mathrm{\Delta}}$
for clusters in superclusters of filament morphology (F, black solid line), 
of spider morphology (S, grey dashed line), and for isolated clusters
(I, thin grey solid line).
}
\label{fig:cummorf}
\end{center}
\end{figure}

Clusters in superclusters of spider morphology have higher probabilities to have 
substructure and larger peculiar velocities of their main galaxies than clusters in 
filament-type superclusters,  being therefore 
dynamically younger (Fig.~\ref{fig:cummorf}). Superclusters of spider morphology 
have richer inner structure than superclusters of filament morphology with large 
number of filaments between clusters in them.  This may lead to the differences 
noted in this study \cite{2012A&A...542A..36E}. 

We note that recently Costa-Duarte et al. 
\cite{2012arXiv1210.0455C} found that the galaxy 
populations in superclusters of filament and pancake type are 
similar. This shows that the problem is not yet solved. Future 
studies of the galaxies and groups in superclusters are needed to 
understand the role of the large scale environment in shaping the properties of 
galaxies and their systems.

\section*{Acknowledgments}
I thank my coauthors Jaan Vennik, Elmo Tempel, Enn Saar, Pasi Nurmi, Antti Ahvensalmi
and others for a fruitful collaboration. 
The present study was supported by the Estonian Science Foundation
grants No. 8005, 7765, 9428, and MJD272, by the Estonian Ministry for Education and
Science research project SF0060067s08, and by the European Structural Funds
grant for the Centre of Excellence "Dark Matter in (Astro)particle Physics and
Cosmology" TK120. 
We are pleased to thank the SDSS Team for the publicly available data
releases. 

\bibliographystyle{ws-procs975x65}
\bibliography{ws-proME}

\begin{thebibliography}{1}

\bibitem{2012A&A...540A.106T}
E.~{Tempel}, E.~{Tago} and L.~J. {Liivam{\"a}gi}, {\em Astron. Astrophys.} {\bf
  540}, p. A106(April 2012).

\bibitem{2012A&A...539A..80L}
L.~J. {Liivam{\"a}gi}, E.~{Tempel} and E.~{Saar}, {\em Astron. Astrophys.} {\bf
  539}, p. A80(March 2012).

\bibitem{2012A&A...540A.123E}
M.~{Einasto}, J.~{Vennik}, P.~{Nurmi}, E.~{Tempel}, A.~{Ahvensalmi}, E.~{Tago},
  L.~J. {Liivam{\"a}gi}, E.~{Saar}, P.~{Hein{\"a}m{\"a}ki}, J.~{Einasto} and
  V.~J. {Mart{\'{\i}}nez}, {\em Astron. Astrophys.} {\bf 540}, p. A123(April
  2012).

\bibitem{2007A&A...476..697E}
M.~{Einasto}, E.~{Saar}, L.~J. {Liivam{\"a}gi}, J.~{Einasto}, E.~{Tago}, V.~J.
  {Mart{\'{\i}}nez}, J.-L. {Starck}, V.~{M{\"u}ller}, P.~{Hein{\"a}m{\"a}ki},
  P.~{Nurmi}, M.~{Gramann} and G.~{H{\"u}tsi}, {\em Astron. Astrophys.} {\bf
  476}, 697(December 2007).

\bibitem{2011ApJ...736...51E}
M.~{Einasto}, L.~J. {Liivam{\"a}gi}, E.~{Tempel}, E.~{Saar}, E.~{Tago},
  P.~{Einasto}, I.~{Enkvist}, J.~{Einasto}, V.~J. {Mart{\'{\i}}nez},
  P.~{Hein{\"a}m{\"a}ki} and P.~{Nurmi}, {\em ApJ} {\bf 736}, p.~51(July 2011).

\bibitem{2012A&A...542A..36E}
M.~{Einasto}, L.~J. {Liivam{\"a}gi}, E.~{Tempel}, E.~{Saar}, J.~{Vennik},
  P.~{Nurmi}, M.~{Gramann}, J.~{Einasto}, E.~{Tago}, P.~{Hein{\"a}m{\"a}ki},
  A.~{Ahvensalmi} and V.~J. {Mart{\'{\i}}nez}, {\em Astron. Astrophys.} {\bf
  542}, p. A36(June 2012).

\bibitem{2012arXiv1210.0455C}
M.~V. {Costa-Duarte}, L.~{Sodre}, Jr and F.~{Durret}, {\em ArXiv: 1210.0455}
  (Octtober 2012).

\end{thebibliography}
\end{document}